\documentclass[12pt,a4paper]{article}

\usepackage{amsmath,amsfonts,amssymb,amsthm,hyperref,cite,authblk,setspace,fullpage,graphicx,color}

\begin{document}

\onehalfspacing

\title{{\small \hfill IFT-UAM/CSIC-14-124, DESY-14-222}\\[1cm]
\textbf{Disentangling the $f(R)$\thinspace -\thinspace Duality}}

\author[1]{Benedict J. Broy}
\author[2]{Francisco G. Pedro}
\author[1]{Alexander Westphal}
\affil[1]{\emph{Deutsches Elektronen-Synchrotron DESY, Theory Group, 22603 Hamburg, Germany}}
\affil[2]{\emph{Departamento de F\'{\i}sica Te\'orica and Instituto de F\'{\i}sica Te\'orica  UAM-CSIC, Universidad Aut\'onoma de Madrid, Cantoblanco, 28049 Madrid, Spain}}

\numberwithin{equation}{section}

\maketitle

\begin{abstract}

Motivated by UV realisations of Starobinsky-like inflation models, we study generic exponential plateau-like potentials to understand whether an exact $f(R)$-formulation may still be obtained when the asymptotic shift-symmetry of the potential is broken for larger field values. Potentials which break the shift symmetry with rising exponentials at large field values only allow for corresponding $f(R)$-descriptions with a leading order term $R^{n}$ with $1<n<2$, regardless of whether the duality is exact or approximate. The $R^2$-term survives as part of a series expansion of the function $f(R)$ and thus cannot maintain a plateau for all field values. We further find a lean and instructive way to obtain a function $f(R)$ describing $m^2\phi^2$-inflation which breaks the shift symmetry with a monomial, and corresponds to effectively logarithmic corrections to an $R+R^2$ model. These examples emphasise that higher order terms in $f(R)$-theory may not be neglected if they are present at all. Additionally, we relate the function $f(R)$ corresponding to chaotic inflation to a more general Jordan frame set-up. In addition, we consider $f(R)$-duals of two given UV examples, both from supergravity and string theory. Finally, we outline the CMB phenomenology of these models which show effects of power suppression at low-$\ell$.

\end{abstract}

\newpage

\tableofcontents

\newpage

\section{Introduction} 

Observational evidence provides increasingly strong support for a very early phase of cosmological inflation. Experiments measuring the cosmic microwave background (CMB)~\cite{9601067,1212.5225, 1303.5082} as well as Large Scale Structure (LSS)  begin to restrict inflation to be driven predominantly by a single scalar field and start to constrain a wide class of potentials. 

Inflation takes quantum fluctuations of light degrees of freedom and stretches them rapidly into super-horizon classical field profiles \cite{Kiefer:1998qe}.  This way, the nearly scale-invariant power spectrum of fluctuations of the inflaton drive variations of the gravitational potential which seed all current large-scale structure and the CMB temperature anisotropy. Moreover, graviton fluctuations produce a nearly scale-invariant spectrum of gravitational waves during inflation which cause B-mode polarization in the CMB. The amplitude of this primordial B-mode spectrum is directly tied to the energy scale of inflation.

With the nature of the degree-scale B-mode signal from BICEP2~\cite{Ade:2014xna} unclear (see e.g.\ \cite{Mortonson:2014bja,Flauger:2014qra,Adam:2014bub}), the current constraints on the fractional gravitational wave power, the tensor-to-scalar ratio,  $r \lesssim 0.1$ allow inflation models with a wide range of energy scales. High-scale models of inflation with tensor mode production at $r\gtrsim 10^{-3}$ include both large-field monomial potentials $V\sim \phi^p$ pioneered by Linde \cite{Linde:1983gd} as well as exponentially flattening-out potentials such as the $R+R^2$ model by Starobinsky \cite{Starobinsky}.

The supergravity embedding of Starobinksy-like inflation models has seen tremendous progress during recent years. Based on studies of non-minimally coupled 'Jordan frame' supergravity~\cite{Ferrara:2010yw,Ketov:2010qz,Ketov:2012jt}, recent works have established various super-conformal realizations of the Starobinsky model in F- and D-term superconformal supergravity models, as discussed in e.g.~\cite{Kallosh:2013lkr,Buchmuller:2013zfa,Kallosh:2013hoa,Ferrara:2013rsa,Farakos:2013cqa,Ketov:2014qoa}. Subsequently this led to the discovery of several rather universal attractor regimes of non-minimally coupled inflation (see e.g.\ \cite{Kallosh:2013tua,Kallosh:2013yoa,Kallosh:2014rga,Kallosh:2014laa}). Models based on no-scale supergravity (see e.g.\ \cite{Ellis2013,Ellis:2014rxa,Ferrara:2014ima,Ellis:2014gxa}) can realize the Starobinsky model in standard F-term supergravity as well.

The Starobinsky~\cite{Starobinsky}  model is appealing both observationally and theoretically, as it may be written in terms of a theory of modified gravity as well as in terms of Einstein gravity with a scalar field  \cite{Whitt84, Vilenkin, Pollock}. Starting from the scalar field description, we will look at the behaviour of this duality between the formulation in terms of a scalar field and the one in terms of an $f(R)$ theory of modified gravity when corrections to the scalar field potential are considered. We will find how different types of corrections to the scalar field potential lead to corrections of logarithmic and power-law type to the $f(R)$-formulation.

Following a quick review of $f(R)$-theory, we show in section \ref{f(R)} that scalar potentials which display an exponentially good restoration of a shift symmetry a large field values $V\sim 1+\sum_n \pm \exp(-c_n\kappa\phi)$ are always dual to an $f(R)$ theory with an asymptotic limit $f(R)\to R^2$.

In section \ref{logR} we demonstrate how potentials with exponentials $\exp(-\gamma\kappa\phi)$ with $0<\gamma<2$ (i.e.\ effectively rescaling the exponential's exponents $\kappa\rightarrow\kappa'$) induce corrections of type $R^{2-\gamma}$ or equivalently (at leading order if $\gamma\ll1$) $\log R$ corrections to the corresponding $f(R)$-dual. This is in line with the observations made in~\cite{1404.3558}, and then~\cite{r2logr,1405.1321,1406.1096,ketov,Motohashi} that adding a term $R^{2-\epsilon}\supset R^2\log R$ can enhance the tensor mode signal significantly over the pure $R+R^2$ case. We show that these models can provide for chaotic inflationary dynamics within the observable range of e-folds. Hence, they yield $f(R)$ duals to chaotic inflation models from logarithmically broken scale invariance in the Jordan frame discussed in \cite{Joergensen:2014rya,Kaloper2014}.

In section \ref{RtoN} we study what happens when we add rising exponential terms to the Einstein frame potential breaking the shift symmetry at large field values. We show that this corresponds to $f(R)$-duals that are to leading order $R^n$ with $1<n<2$. For certain setups, we can give a closed form expression. Hence, in these cases we can understand the rising exponentials in the Einstein frame as an infinite series of higher scalar-curvature corrections $\sum_p c_p R^p$ to the Einstein-Hilbert action. 

Section \ref{RtoNuv} discusses toy models of UV completion both in the context of supergravity and type IIB string theory for setups of the previous section. We derive approximate and exact $f(R)$-duals for the exemplary low energy effective potentials.

Finally,  in section~\ref{LowEll} we apply our models with rising exponential terms to some of the aspects of the  suppression of CMB power at large angular scales. Exponential terms rising out of the $R^2$-dominated plateau provide a source of rapid steepening of the scalar potential in the Einstein frame. Hence, they limit the total amount of slow-roll e-folds and the built-in steepening is a mechanism for the suppression of CMB power at large angular scales studied in recent literature~\cite{0303636,1211.1707,1309.3412,1309.3413,1309.4060,1404.2278,Kallosh:2014xwa}. Moreover, we find that the corrections to the scalar field potential required to have a strong suppression effect on large-angle CMB power no longer have an exact but only an asymptotic $f(R)$-dual. 
We discuss our results in section~\ref{concl}.

\section{$f(R)\to R^2$ -- Exponential Restoration of the Shift Symmetry}\label{f(R)}

We start this section with a brief review of the essentials of $f(R)$-theory (we refer the reader looking for a more thorough review to  e.g.\ \cite{lrr-2010-3}). Consider a modified Einstein-Hilbert Lagrangian of the form
\begin{equation}\label{Lagrangian}
\mathcal L=\frac{1}{2}\sqrt{-g}f(R),
\end{equation}
with the Ricci scalar and tensor given by $R=g^{\mu\nu}R_{\mu\nu}$ and $R_{\mu\nu}=R^\alpha_{\mu\alpha\nu}$ respectively. By performing a conformal transformation on the metric $g_{\mu\nu}$ with conformal factor $\Omega$:
\begin{equation}\label{transformation}
\tilde g_{\mu\nu}=\Omega\  g_{\mu\nu}, \quad \sqrt{-\tilde g}=\Omega^{2}\ \sqrt{- g},
\end{equation}
one can show that the Ricci scalar transforms as
\begin{equation}\label{Ricci}
R=\Omega\left(\tilde R + 6\tilde\Box \omega -6\tilde g^{\mu\nu}\partial_\mu\omega\partial_\nu\omega\right),
\end{equation}
with $\omega\equiv (1/2)\ln\Omega$ and $\tilde\Box=\tilde\nabla^\mu\tilde\nabla_\mu$. The behaviour of the Ricci scalar under this Weyl transformation makes apparent the additional degree of freedom, $\omega$, present in Eq. (\ref{Lagrangian}). Let us now rewrite the Jordan frame Lagrangian \eqref{Lagrangian} as
\begin{equation}\label{RewrittenL}
\mathcal L=\sqrt{-g}\left(\frac{1}{2}f' R - U \right),
\end{equation}
where the prime denotes differentiation with respect to $R$ and we have introduced $U=(1/2)(f'R-f)$. Substituting \eqref{Ricci} into \eqref{RewrittenL} and applying \eqref{transformation} yields
\begin{equation}\label{Weyl}
\mathcal L=\sqrt{-\tilde g}\left[\frac{1}{2}\Omega^{-1}f'\left(\tilde R-6\tilde g^{\mu\nu}\partial_\mu\omega\partial_\nu\omega \right)-\Omega^{-2} U\right],
\end{equation}
where we have omitted the $\tilde\Box\omega$ term as its contribution to the action vanishes due to Gauss' theorem. Defining $\phi\equiv \sqrt{3/2}\ln\Omega$ normalises the kinetic term\footnote{We could also have chosen $\phi=-\sqrt{3/2}\ln\Omega$. This essentially mirrors the resulting potential with $V(\phi)\rightarrow V(-\phi)$.}. Choosing the conformal factor $\Omega=f'$ for $f'>0$ brings the action to the Einstein frame where the Lagrangian takes the form
\begin{equation}\label{finalL}
\mathcal L=\sqrt{-\tilde g}\left[\frac{\tilde R}{2}-\frac{1}{2}\tilde g^{\mu\nu}\partial_\mu\phi\partial_\nu\phi -V(\phi)\right],
\end{equation}
with the potential for the scalar degree of freedom given by
\begin{equation}
V(\phi)=(f'R-f)/(2f'^2).
\label{potentialDef}
\end{equation}
The relation between the Jordan frame scalar curvature and the canonically normalised scalar field, $R(\phi)$, can be found by inverting $\phi=\sqrt{3/2}\ln f'$ for a given $f(R)$. Thus an $f(R)$-theory may be recast in terms of gravity with a scalar field, provided $f'$ is invertible to allow for a relation between the Ricci scalar and the canonically normalised field $\phi$. 

\subsection*{Examples}

In this work we will make extensive use of the mapping between the Jordan frame $f(R)$ and the Einstein frame scalar potential $V(\phi)$. With that in mind we now demonstrate how to obtain a potential $V(\phi)$ from a given $f(R)$ and vice-versa for the well known Starobinsky model \cite{Pollock}
\begin{equation}
\label{starof}
f(R)=R+\alpha R^2.
\end{equation}
Choosing $f'=e^{\kappa\phi}$, we find $R(\phi)=(2\alpha)^{-1}(e^{\kappa\phi}-1)$. Evaluating the expression for $V(\phi)$ of \eqref{finalL} then yields
\begin{equation}\label{staro}
V(\phi)=V_0\left(1-e^{-\kappa\phi} \right)^2,
\end{equation}
where $\kappa=\sqrt{2/3}$ and $V_0=1/8\alpha$. This is the well know Starobinsky potential that provides a model for cosmic inflation along the plateau of the potential, where $V\sim V_0$.

Let us consider a small deformation of the theory (\ref{staro}) and contemplate a potential of the form 
\begin{equation}\label{fibre}
V(\phi)=V_0\left(1-C_0 e^{-\frac{\kappa}{2}\phi}+C_1 e^{-2\kappa\phi}\right),
\end{equation}
with $C_0-C_1=1$ to ensure $V(0)=0$. The difference to potential \eqref{staro} is that the second exponential is no longer the square of the first. Even though this does not significantly affect the inflationary phenomenology, the nature of the duality changes as we will show in the following. Identifying $f'=e^{\kappa\phi}$, we recall the relation between Einstein frame's $V(\phi)$ and Jordan frame's $f(R)$, Eq. (\ref{potentialDef}),  and thus consider the differential equation
\begin{equation}\label{boundary}
\frac{f'R-f}{2f'^2}=V_0\left(1-\frac{C_0}{\sqrt{f'}}+\frac{C_1}{f'^2} \right).
\end{equation}
By solving this equation one may find the $f(R)$ theory that is dual to the potential of Eq. (\ref{fibre}). Multiplying the above with $2f'^2$ and differentiating with respect to either $R$, $f$, or $f'$ gives
\begin{equation}
f'-\frac{3}{4}C_0\sqrt{f'}-\frac{1}{4V_0}R=0,
\end{equation}
provided $f''\neq0$. This can be solved as
\begin{equation}\label{sign}
f'=2\left(\frac{3}{8}C_0\right)^2+\frac{1}{4V_0}R\pm\frac{3}{4}C_0\sqrt{\left(\frac{3}{8}C_0\right)^2+\frac{1}{4V_0}R}.
\end{equation}
Considering \eqref{boundary} as a boundary condition\footnote{{I.e.\ the integration constant is chosen such that the resulting function $f(R)$ satisfies \eqref{boundary} when plugged back in. Thus the information which has been lost upon differentiating \eqref{boundary} is regained.}}, we can then integrate the above to find the corresponding $f(R)$-theory to be
\begin{equation}
f(R)=\frac{R}{2}+\frac{8V_0}{3}\left(\frac{1}{4}+\frac{R}{4V_0}\right)^{3/2}+\frac{R^2}{8V_0}-\frac{V_0}{3},
\end{equation}
where we have chosen $C_0=4/3$ and $C_1=1/3$ such that $V(0)=0$ and the positive sign in \eqref{sign} so that $f(R)\geq 0$ and $f(0)=0$. Note that the above is $\sim R^2$ in the large $R$ regime but has a modified behaviour for intermediate $R$. This comes as no surprise since potential \eqref{fibre} also has an approximate shift-symmetry for large field values but, as opposed to \eqref{staro}, does not display the quadratic relation among the potential's exponentials which in turn influences the behaviour of the theory at intermediate $R$ and $\phi$. 

In principle, any potential $V=V_0(1-2e^{-\frac{\kappa}{n}\phi}+e^{-2\kappa\phi})$ with $n>1$, which hence breaks the square relation of the two exponentials, has an approximate dual $f(R)\sim R^2$ for large $R$. Similarly, breaking the square relation with the second exponential such that $V=V_0(1-2e^{-\kappa\phi}+e^{-n\kappa\phi})$ with $n>2$ also has $f(R)\sim R^2$ as the large $R$ dual when the corresponding differential equation may not be solved explicitly any more\footnote{{We would like to refer the reader to appendix \ref{a1} for a concise demonstration of these statements.}}. Both results are expected as the corresponding potentials maintain an approximate shift-symmetry at large field values. This can further be demonstrated by simply considering $f(R)=\alpha R^2$ which reduces to a pure cosmological constant $\Lambda=(8\alpha)^{-1}$ in the Einstein frame.

\section{Restoration at Large Fields -- Logarithmic Corrections to $f(R)$}\label{logR}

\subsection{Changing the Coefficient $\kappa$}

The line of argument given above requires\footnote{Applying the above procedure with $f'=e^{\kappa'\phi}$, where $\kappa'\neq\sqrt{2/3}$, introduces a non-canonical kinetic term in \eqref{finalL}.} the potential to contain exponentials with a coefficient $\kappa=\sqrt{2/3}$ in the exponent. However, in the same spirit that led us to deviate from the square relation between the exponentials in  $V$, we may now consider potential \eqref{staro}  {with a change of the exponent's coefficient, i.e.} with $\kappa\rightarrow\kappa'\neq\kappa$:
\begin{equation}\label{threepointone}
V(\phi)=V_0\left(1-2 e^{-\kappa'\phi}+ e^{-2\kappa'\phi} \right),
\end{equation}
and try to understand to what extent this deformation of the original model alters the gravitational $f(R)$ description.

To apply the previous method, we have to recast potential {\eqref{threepointone} with coefficient $\kappa '$} in terms of $\kappa=\sqrt{2/3}$. When {we choose} e.g.\ $\kappa'=2/\sqrt{3}{\thinspace=\sqrt{2}\thinspace \kappa}$, we {write \eqref{threepointone}} in terms of $\kappa$ {as}
\begin{equation}
V(\phi)=V_0\left(1-2 e^{-\sqrt{2}\kappa\phi}+ e^{-2\sqrt{2}\kappa\phi} \right).
\end{equation}
Rather than attempting to analytically solve a differential equation with irrational powers in $f'$, we now integrate implicitly to find
\begin{equation}
f(R)=\frac{R^2}{8V_0}+\alpha\int f'^{1-\sqrt{2}}dR-\beta\int f'^{1-2\sqrt{2}}dR,
\end{equation} 
where $\alpha, \beta$ depend on $V_0$ and the $\phi$ - coefficients of the exponentials in {\eqref{threepointone}}. Considering the inflationary regime, i.e.\ imposing the regime of large $R$, we approximate $f'\sim R$ and write
\begin{equation}\label{chaoticf}
f(R)=\frac{1}{8V_0}R^2+c \ R^{2-\sqrt{2}}+\ldots
\end{equation}
where $c$ is a rescaled $\alpha$ due to the integration. Hence a given inflaton potential with a rescaled exponent such as 
\begin{equation}\label{chaotic}
V(\phi)= V_0\left(1-e^{-\gamma \sqrt{\frac{2}{3}}\phi} \right)^2,
\end{equation}
where $0<\gamma<2$, has an approximate $f(R)$ dual given by
\begin{equation}\label{larger}
f(R)=\frac{1}{8V_0}R^2+cR^{2-\gamma}+\ldots
\end{equation}
up to sub-leading terms during slow-roll inflation {(see appendix \ref{a2} for an explicit argument)}.

\subsection{Chaotic Inflation from $f(R)$-theory}

Observable inflation in the Starobinsky model proceeds in the transition region between the flat plateau and the minimum of the potential, ending at $\phi\sim 1$ $M_{Pl}$. Observationally, it yields a spectral index compatible with observations and very low primordial tensor fluctuations. The phenomenologist may ask if there is a deformation of the standard Starobinsky model that significantly changes the observational signatures, making them more in line with those of chaotic inflation and if so what is its corresponding gravitational dual.
\begin{figure}[htp]
\centering
\includegraphics[scale=0.7]{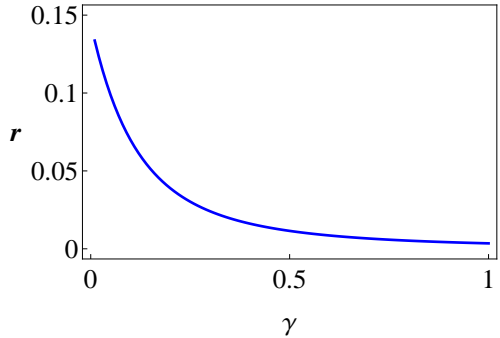}
\caption{\emph{{The plot depicts the observable tensor to scalar ratio $r$ for potential \eqref{chaotic} or equivalently its dual \eqref{larger} as a function of the parameter $\gamma$ in the range $0<\gamma<1$. For sufficiently small $\gamma$, the value of $r$ converges to the prediction of $m^2\phi^2$ inflation. The tensor to scalar ratio has been calculated 55 e-folds before the end of inflation.}}}
\label{potential}
\end{figure}

Note that potential \eqref{chaotic} may well approximate chaotic inflation for intermediate field values provided $\gamma\ll 1$. {To see this, consider the series expansion of \eqref{chaotic} around the point $\phi=0$
\begin{equation}
V(\phi)\approx\kappa^2\gamma^2\phi^2-\kappa^3\gamma^3\phi^3+\mathcal O(\gamma^4\phi^4).
\end{equation}
Hence for $\gamma\ll 1$, inflation occurs in the concave region of the potential, i.e.\ before the nearly shift-symmetric plateau.} Thus \eqref{larger} can be an $f(R)$ dual for a potential that has a chaotic regime before approaching the shift-symmetric plateau. Hence adding a term of type $R^{2-\gamma}$ with $\gamma\ll 1$ to a Starobinsky $f(R)$ can induce a chaotic regime and hence change the predictions for the spectral index $n_s$ and tensor-to-scalar ratio $r$ to those of $\phi^m$ models, depending on the specific value of $\gamma$ {as depicted in figure 1}. This confirms the findings of \cite{Sebastiani:2013eqa,1404.3558,1406.1096,1405.1321,Motohashi} but without omitting the $R^2$-term in the first place. The $R^2$-induced plateau can simply be pushed beyond the last 60 e-foldings of inflation and hence does not influence the inflationary predictions, yet it still determines the behaviour at large field values or e-foldings corresponding to super-horizon scales. 

Now consider $\gamma\ll 1$ and expand \eqref{larger} as
\begin{align}
 f(R)&\approx R^2+R^{2-\gamma}+\ldots \notag \\
 &\approx R^2\left[1+e^{-\gamma\ln R} \right] +\ldots \notag \\
 &\approx R^2\left[1+\sum_{n=0}^\infty\frac{\left(-\gamma\ln R \right)^n}{n!}  \right]+\ldots \notag \\
 &\approx R^2 -\gamma R^2\ln R +\frac{\gamma^2}{2}  R^2 \ln^2 R -\mathcal O(\gamma^3\ln^3 R) +\ldots \label{logr}
\end{align}
The first two terms of \eqref{logr} resemble the ansatz of \cite{r2logr}. Revisiting Starobinsky's original idea,  it was investigated whether or not logarithmic corrections to $R^2$-inflation change the inflationary observables, specifically whether a larger amount of tensor modes is produced. It was found that even though a first-order logarithmic correction lifts the potential's plateau, a significant enhancement of tensor modes does not occur. 

We hence realise that although higher order terms seem to be sub-leading at first sight, they change the inflationary dynamics drastically. Whereas a linear logarithmic correction only causes a slight deviation in the inflationary observables, higher-order terms can indeed make the inflationary dynamics chaotic within the accessible range of inflationary e-folds. Thus if one assumes higher-order terms to exist, they have to be considered. This confirms the findings of \cite{ketov}, where parametric methods are used to obtain an $f(R)$-theory corresponding to $m^2\phi^2$-inflation and it is found that a linear logarithmic correction is not sufficient but a squared term is necessary. 

\subsection{Another Jordan Frame}

We have seen above how to obtain a function $f(R)$ describing the dynamics of chaotic inflation. However, one may also ask how a corresponding scalar field theory being non-minimally coupled to gravity looks like, hoping to learn something more general regarding the mapping between some Jordan and the Einstein frame. 

One can always go from the Starobinsky frame to the Einstein frame and then try to find another Jordan frame again. We now propose how to go from the Starobinsky frame to a specific Jordan frame directly, namely the Jordan frame with a non-minimal coupling function
\begin{equation}\label{coupling}
\varOmega (\phi)=1+\xi\phi^2.
\end{equation}
For the remainder of this section we will use $\tilde R$ to denote the Starobinsky frame's curvature scalar, while $R$ denotes scalar curvature in the new Jordan frame. We start by first recalling a slightly rewritten Lagrangian \eqref{Weyl}, i.e.\ the Weyl-transformed Lagrangian of some $f(\tilde R)$-theory, 
\begin{equation}\label{Weyl2}
\frac{\mathcal L}{\sqrt{-g}}=\Omega^{-1}f'\left[\frac{R}{2}-\frac{3}{4}\left(\partial\log\Omega \right)^2 \right]-\frac{U^2}{\Omega^2},
\end{equation}
where we have rescaled the metric with the function $\Omega$. For \eqref{Weyl2} to simply reduce to the Einstein frame, we would just choose $\Omega=f'$.

However, now we seek a Jordan frame given by \eqref{coupling}, hence in \eqref{Weyl2} we require
\begin{equation}\label{label1}
\Omega^{-1}f'\equiv 1+\xi\phi^2.
\end{equation}
We then have to normalise the kinetic term accordingly in order to obtain a relation between $\Omega$ and $\phi$. We thus write
\begin{equation}
\frac{3}{4}\left(1+\xi\phi^2 \right)\left(\partial\log\Omega \right)^2\equiv\frac{1}{2}\left(\partial\phi \right)^2,
\end{equation}
which, in the regime of strong coupling, i.e.\ $\xi\gg 1$, may be solved to yield
\begin{equation}
\Omega(\phi)=\phi^{\Delta},
\end{equation}
where $\Delta=\kappa/\sqrt{\xi}$ with $\kappa=\sqrt{2/3}$. Thus for strong coupling, $\xi\gg1{\Leftrightarrow} \Delta\ll 1$. Combining the above with \eqref{label1} gives
\begin{equation}\label{RelationRPhi}
f'=\phi^{\Delta}\left(1+\xi\phi^2 \right).
\end{equation}
The resulting Jordan frame potential can then be written as
\begin{align}\label{JPot}
V_J(\phi)&=\frac{\phi^{-2\Delta}}{2}\left(f'\tilde R-f \right),
\end{align}
where $\tilde R$ denotes the Starobinsky frame's curvature scalar. Then $\tilde R=\tilde R(\phi)$ as well as $f(\tilde R(\phi))$ can be obtained from specifying a certain $f(\tilde R)$ and obtaining the relation between $\tilde R$ and $\phi$ from \eqref{RelationRPhi}. The resulting Jordan frame Lagrangian is then given as
\begin{equation}
\frac{{\mathcal L}_J}{\sqrt{-g}}=\left(1+\xi\phi^2\right)\frac{R}{2}-\frac{1}{2}\left(\partial\phi \right)^2-\frac{\phi^{-2\Delta}}{2}\left[f'(\tilde R(\phi)) \tilde R(\phi)-f(\tilde R(\phi)) \right].
\end{equation}

In case of the Starobinsky model $\tilde R+\alpha \tilde R^2$, one indeed recovers a non-minimally coupled scalar field theory with a $\phi^4$ potential. One finds
\begin{equation}
\tilde R(\phi)\sim\phi^{2+\Delta}
\end{equation}
and therefore
\begin{equation}
V_J(\phi)\sim\frac{\phi^{-2\Delta}}{2}\tilde R(\phi)^2\rightarrow \phi^4,
\end{equation}
which is the leading-order potential of the well-known non-minimally coupled $\phi^4$-inflation model \cite{Salopek,0710.3755}

Let us now investigate, whether we can directly find a corresponding non-minimally coupled scalar field theory that is dual to an inflationary period driven by a term $\tilde R^{2-\gamma}$ with $\gamma\ll 1$. To that end, consider 
\begin{equation}
f'(\tilde R)=(2-\gamma)\tilde R^{1-\gamma}
\end{equation}
as the derivative of the function $f(\tilde R)$ for the range of observable e-folds. The potential \eqref{JPot} then is
\begin{equation}\label{Deformed}
V_J(\phi)=\frac{\phi^{-2\Delta}}{2}\left[\tilde R(\phi)^{2-\gamma}-\gamma \tilde R(\phi)^{2-\gamma} \right],
\end{equation}
where the second term in the brackets is sub-leading as $\gamma\ll 1$. We further have from \eqref{RelationRPhi} at large $\xi$
\begin{equation}
f'\sim \phi^{2+\Delta}
\end{equation}
and thus
\begin{equation}
\tilde R(\phi)\sim \phi^{(2+\Delta)/(1-\gamma)}.
\end{equation}
Plugging the above back into \eqref{Deformed} gives, upon expansion of the exponent
\begin{equation}
V_J(\phi)\sim\phi^4\phi^{2\gamma+\mathcal O(\gamma^2)}.
\end{equation}
This can then be rewritten as
\begin{equation}
V_J(\phi)\sim\phi^4\left[1+c_1\gamma\log\phi+c_2\gamma^2\log^2\phi+\mathcal O(\gamma^3) \right]
\end{equation}
and hence may be understood as a quantum corrected non-minimally coupled model of $\phi^4$ inflation\footnote{These results may easily be extended to models in which the non-minimal coupling scales as $\varOmega(\phi)\sim \sqrt{V(\phi)}$ such as \cite{1310.3950}, given the frame function is to leading order of power law type.}.

This demonstrates how logarithmic corrections to some function $f(R)$ can be directly related to logarithmic corrections to the potential in another Jordan frame. Hence our results here close the $f(R)\to V_J\to V_E\to f(R)$ circle started by the analysis in \cite{Kaloper2014} where such logarithmic corrections to a $\phi^4$ Jordan frame potential were found to produce Einstein frame quadratic large-field inflation\footnote{This can nicely be demonstrated by considering a Jordan frame Lagrangian $\mathcal L_J=\sqrt{-g}\left[\Omega(\phi)R/2-1/2\thinspace (\partial\phi)^2 -V_J\right]$ with $V_J\propto[\Omega(\phi)\log\Omega(\phi)]^2$ for arbitrary frame function $\Omega(\phi)$. This is an immediate generalization of the observation made in~\cite{Kaloper2014}. In the Einstein frame, this theory has a purely quadratic potential for a canonically normalised inflaton.}.

\section{Exponential Shift Symmetry Breaking -- Power-Law Corrections to $f(R)$}\label{RtoN}

\subsection{Rising Exponentials}

Returning to Starobinsky-esque potentials, we now want to determine functions $f(R)$ corresponding to a class of potentials which break the shift-symmetry of the inflationary plateau by a rising exponential at large field values. Hence consider e.g.
\begin{equation}\label{V2}
V(\phi)= V_0\left(C_{3}-C_0 e^{-\frac{\kappa}{2}\phi}+C_1 e^{-2\kappa\phi}+C_2 e^{\kappa\phi} \right),
\end{equation}
where $C_2\ll C_{0},C_1, C_3$ and the sum over all $C_i$ is unity. We recall the expression for $V(\phi)$ in terms of $f(R)$ and its derivative, Eq. (\ref{potentialDef}) and obtain the differential equation
\begin{equation}\label{harddiff}
f'^2+ \frac{2 C_3}{3 C_2}f'-\frac{C_0}{2C_2}\sqrt{f'}-\frac{1}{6 V_0 C_2}R=0.
\end{equation}
As in the previous section one would like to solve this equation to determine the exact $f(R)$ dual to the potential of Eq. (\ref{V2}), however (\ref{harddiff}) above is not easily solvable. We hence learn that adding a rising exponential to the potential \eqref{fibre} prevents us from finding a dual $f(R)$ description for the entire field range. Nevertheless asymptotic results are still within reach. Focusing just on the rising exponential, we may now want to find the approximate $f(R)$-dual for potentials that are generically
\begin{equation}
V(\phi)\sim V_0\ e^{n\kappa\phi},
\end{equation}
at large field values, with $n\geq 1$. {Recalling \eqref{potentialDef}, we write
\begin{equation}
V_0f'^n=\frac{f'R-f}{2f'^2},
\end{equation}
which we rearrange as
\begin{equation}
2V_0f'^{n+2}-f'R+f=0.
\end{equation}
Differentiating with respect to $R$ and solving for $f'$ then yields
\begin{equation}
f'=\frac{1}{2(n+2)V_0 }R^{1/(n+1)}.
\end{equation}
Integrating, we find the leading order solution}
\begin{equation}\label{genericn}
f(R)\sim R^{(n+2)/(n+1)}.
\end{equation}
Thus even though the full analytic expression cannot always be found, it is still possible to give the approximate form for a high-energy $f(R)$-theory, where the resulting $f(R)$ is always $R^n$ with $n<2$ to leading order. 

Considering a correction with very large $n$, we see that the resulting $f(R)$ corresponds to a theory where the exponent of the non-linear term may effectively be understood as a perturbation around unity. Similarly, we saw above that a term $R^{2-\gamma}$ with $\gamma\ll 1$ produces a chaotic regime, with the specific inflationary signatures depending on the tuning of $\gamma$. This can be understood, as has already been pointed out in \cite{Chakravarty:2013eqa,1404.3558,1406.1096,1405.1321}, as a correction of an $R^2$ term around its exponent. Thus we see that small deviations from integer powers in a polynomial function $f(R)$ have a major effect on the resulting potential and hence the inflationary observables. 

\subsection{Maintaining a Plateau}

The fact that a $R^n$ term with $n>2$ cannot steepen the corresponding potential could also have been foreseen from the following argument. Consider a function $f(R)$ of the form
\begin{equation}
f(R)=\sum_{n=1}^{N}a_n R^n
\end{equation}
where all $a_n>0$. In the limit of large $R$, $f(R)\rightarrow a_N R^N$ and hence $f'\sim R^{N-1}$. We now seek the behaviour of the potential for very large $\phi$ and thus very large $R$. Considering expression \eqref{potentialDef} for $V(\phi)$, we obtain
\begin{equation}
V(\phi(R))\rightarrow R^{2-N}.
\end{equation}
Hence it appears that only $N=2$ produces a plateau in the potential at large $\phi$, any power $N<2$ curves the potential upwards whereas powers of $N>2$ have it asymptote the axis. This is in line with the early work of \cite{Maeda:1988}.

\subsection{Finite Order Corrections}

We now consider whether there is a way to steepen the potential by considering finite higher order corrections to an $R+R^2$  theory, where the coefficients of higher order terms may be negative. As we established above, terms with positive coefficients and power $n>2$ curve the potential downwards towards the axis. Thus consider a function $f(R)$ such that
\begin{equation}\label{rcubed}
f(R)= R+\alpha R^2 + \beta R^3,
\end{equation}
but with $\beta<0$. We then find
\begin{equation}\label{toInvert}
\phi=\sqrt{3/2}\ln\left(1+2\alpha R -3|\beta|R^2 \right),
\end{equation}
or equivalently
\begin{equation}
R(\phi)=\frac{\alpha}{3|\beta|}\pm\sqrt{\frac{1}{9|\beta|^2}+\frac{1}{3|\beta|}-e^{\kappa\phi}}.
\end{equation}
Thus the field $\phi$ is only defined as long as $f'>0$, in other words, the field space is limited\footnote{This modification has already been investigated in \cite{BandM}. Importantly, it was argued that a universe with $f'=0$ for some $R$ consists of two causally disconnected regions, one in which $f'<0$ and one with $f'>0$. Hence a universe with $f'<0$ can never evolve into Minkowski space.}. Equivalently, the Ricci scalar $R$ becomes complex when $f'$ changes sign. For $f'\rightarrow 0$, $\phi\rightarrow -\infty$. This demonstrates that the spacetime region where $f'<0$ is disconnected from the $f'>0$ region as it is pushed infinitely far away in field space. Further, it is known (see e.g.\ \cite{lrr-2010-3}) that ghosts appear once $f'<0$. However, a negative coefficient $\beta$ will introduce some steepening over a finite range. 

We further note that higher powers in $R$ with negative coefficients do not increase the steepening of the potential but simply shorten the field range over which the field $\phi$ is defined{, as depicted in the first plot of figure 2}. Again, higher powers are only tractable as long as \eqref{toInvert} remains analytically soluble for $R$.

\subsection{A full $f(R)$-Toy Model}

Having established result \eqref{genericn}, we may now however ask if there exists a function $f(R)$ that is not dominated by a quadratic term at large $R$, but - in its series expansion - has a quadratic term that dominates in an intermediate regime. 

In the Einstein frame language this corresponds to having a scalar field potential that has a plateau separating the minimum at the origin from a growing exponential at large field values (see the right plot of figure 2). Consider
\begin{equation}\label{risingStar}
V(\phi)=V_0\left(1-2e^{-\kappa\phi}+e^{-2\kappa\phi}+\delta e^{n\kappa\phi} \right)-\delta\  V_0,
\end{equation}
with $\delta\ll 1$ and $n=1$. The subtraction of $\delta \ V_0$ ensures $V(0)=0$. Following the usual method we find the differential equation
\begin{equation}\label{differentialequation}
f'^2+\frac{2}{3\ \delta}(1-\delta)f'-\frac{2}{3\ \delta}-\frac{R}{6\ \delta\ V_0}=0,
\end{equation}
from which we obtain an equation for $R$, namely
\begin{equation}\label{expressionForR}
R=6V_0\delta  f'^2 +4V_0(1-\delta)f'-4V_0.
\end{equation}
Substituting $f'=e^{\kappa\phi}$ then yields $R(\phi)$ for the potential of Eq. \eqref{risingStar}. The solution to \eqref{differentialequation} reads
\begin{equation}\label{risingF}
f(R)=\frac{\delta-1}{3\delta}R+4\delta V_0\left[\frac{(1-\delta)^2}{9\delta^2}+\frac{2}{3\delta}+\frac{R}{6\delta V_0} \right]^{3/2}+K,
\end{equation}
where $K$ is a constant of integration determined by a boundary condition as in \eqref{boundary}. This constitutes one of the main results of this work. Explicitly, {the integration constant} $K$ has to be chosen such that the function $f(R)$ satisfies {the boundary condition \eqref{boundary}}
\begin{equation}
\frac{f'\cdot R - f}{2f'^2}=V_0\left(1-2e^{-\kappa\phi}+e^{-2\kappa\phi}+\delta e^{n\kappa\phi} \right)-\delta\cdot V_0,
\end{equation}
where again $e^{\kappa\phi}=f'$. {Satisfying \eqref{boundary} also ensures that $f(0)=0$ and hence a Taylor expansion of \eqref{risingF} recovers the Starobinsky model to leading order.} Thus adding a rising exponential to the Starobinsky potential \eqref{staro} yields a dual $f(R)$ description \eqref{risingF} that is not dominated by a quadratic term at large $R$. Note that the above holds for $n=1$ in the rising exponential. A higher $n$ increases the leading power in \eqref{differentialequation} and induces logarithmic and inverse trigonometric terms in the $f(R)$ description. Analytic solutions of \eqref{differentialequation} in terms of $f'$ only exist up to some $n$. We hence focus on the case $n=1$. 

Further note that for the above, we find from \eqref{expressionForR} 
\begin{equation}
	R|_{\phi=0}=2\delta V_0,
\end{equation}
which is $\ll\mathcal O(10^{-10})$ in Planck units. Thus there is a lower bound on $R$ when $\phi=0$. We further find that
\begin{equation}
	f(R|_{\phi=0})=2\delta V_0,
\end{equation}
and, as expected, $f(0)=0$. Hence when the field $\phi$ has reached its minimum at $V(0)=0$, the corresponding $f(R)$-description displays a cosmological constant type term. This is also indicated by the fact that $f'|_{R=0}<1$ but $f'|_{R=2\delta V_0}=1$. Importantly, switching off the perturbation by sending $\delta\to 0$ restores $R|_{\phi=0}=0$ and thus the Starobinsky model. We may further demonstrate this limiting behaviour by Taylor expanding \eqref{risingF} to find the coefficient of the linear and the quadratic term in $R$ to approach the value of the coefficients of the Starobinsky model when $\delta\rightarrow 0$. 

Let us stress that \eqref{risingF} is an $f(R)$ theory with an infinite number of terms in its series expansion where the term $c_2 R^2$ induces the inflationary plateau. The infinite number of higher power terms does not curve the potential down but sums to an contribution that is to leading order $R^{3/2}$ and hence steepens the potential. In an EFT sense, \eqref{risingF} gives the full theory and determines the coefficients of every power. 
\begin{figure}[htp]
\centering
\includegraphics[scale=0.654]{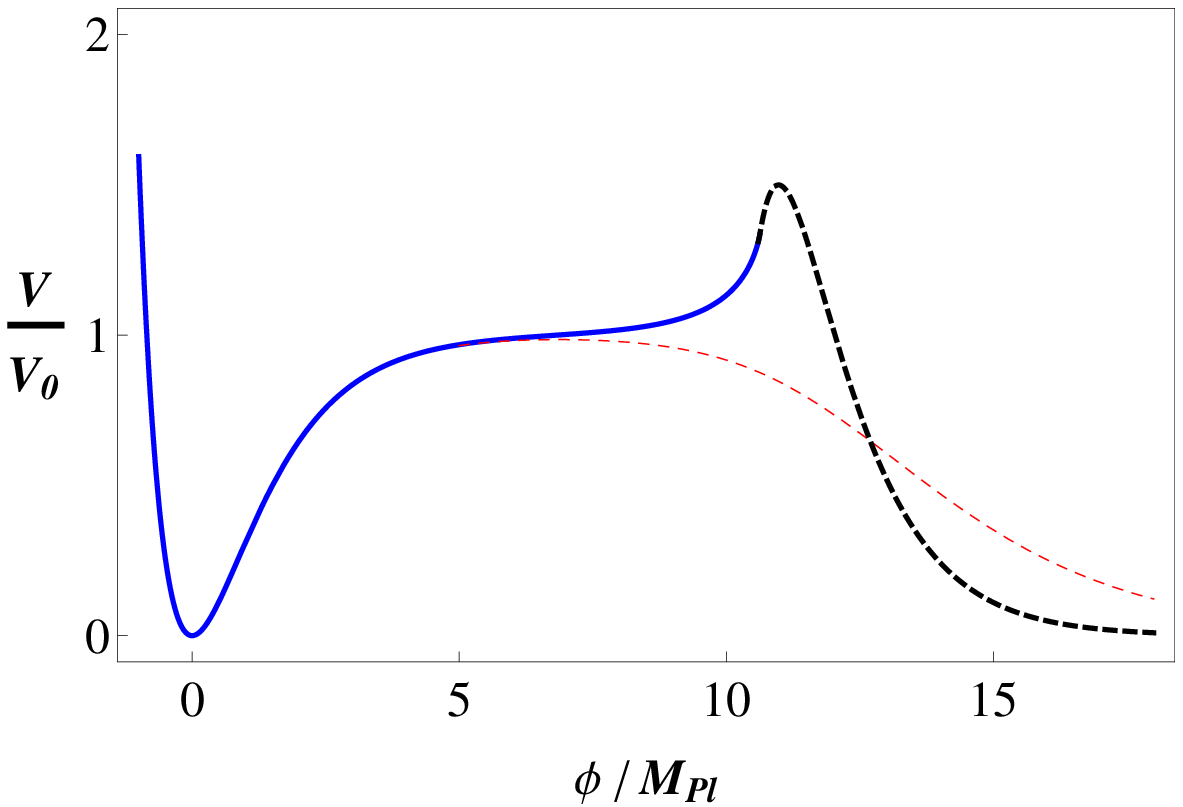}
\includegraphics[scale=0.654]{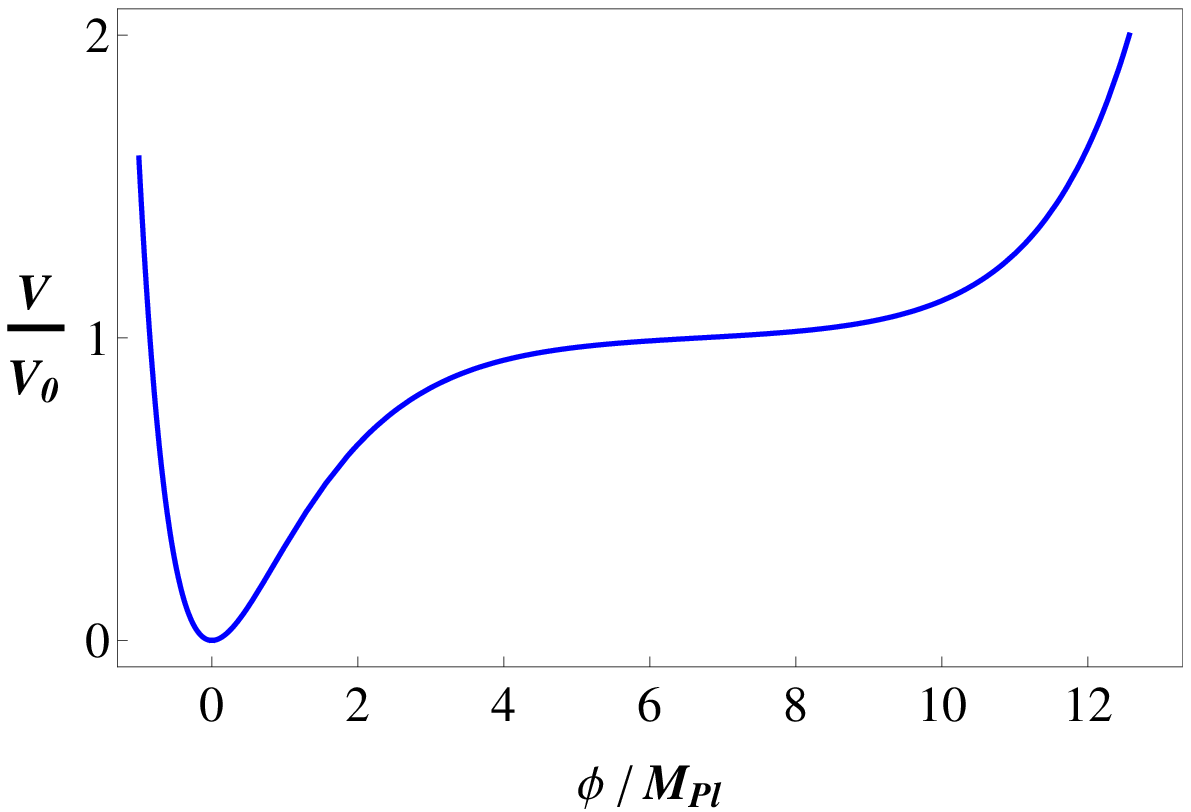}
\caption{\textbf{Left:} \emph{The blue line depicts the potential dual to \eqref{rcubed}. The black dashed line is the real continuation of the potential once $R$ has become complex. The red dashed line displays the potential dual to a finite $f(R)$-theory, where the highest order term is $\sim R^3$ and has a positive coefficient.} \textbf{Right:} \emph{The Einstein frame potential dual to the $f(R)$-theory \eqref{risingF}. Contrary to \eqref{rcubed}, the field range is not limited, but the potential is lifted infinitely after the $c_2 R^2$-term induced and intermediate plateau. In both cases, we have normalised the potential such that $V_0\sim 10^{-10}$ in Planck units.}}
\label{potential}
\end{figure}

\section{UV Examples of Exponential Shift Symmetry Breaking}\label{RtoNuv}

We now turn our attention to examples of inflationary potentials with intermediate shift-symmetry that can be embedded into  a candidate UV theory.

\subsection{No-Scale Supergravity}

Let us start by considering the scenario of \cite{Ellis2013}, where the inflaton superfield is described by a Wess-Zumino model. The potential for the real part of the inflaton superfield reads
\begin{equation}\label{wess}
V(\phi)=\mu^2 \sinh^2 \left(\frac{\phi}{\sqrt{6}} \right)\left[\cosh \left(\frac{\phi}{\sqrt{6}} \right) - \frac{3\lambda}{\mu}\sinh \left(\frac{\phi}{\sqrt{6}} \right) \right]^2,
\end{equation}
with $\phi$ driving inflation and $\mu, \lambda$ being parameters of the model. By expanding the hyperbolic functions the potential \eqref{wess} can be written in terms of exponentials as 
\begin{equation}\label{coefficients}
V(\phi)=C_0 e^{2\kappa\phi}+C_1 e^{\kappa\phi}+C_2 e^{-\kappa\phi}+C_3 e^{-2\kappa\phi}+C_4,
\end{equation}
where the $C_i$ are dependent on $\mu, \lambda$, the sum over all $C_i$ is zero and $\kappa=\sqrt{2/3}$. Judiciously choosing the coefficients of the quadratic and cubic terms in the Wess-Zumino superpotential such that $\lambda/\mu=1/3$, the coefficients $C_0, C_1$ vanish and the above reduces to 
\begin{equation}
V(\phi)=\mu^2 e^{-\sqrt{2/3}\phi}\sinh^2\left(\frac{\phi}{\sqrt{6}} \right) ,
\end{equation}
which is the Starobinsky potential and hence has $R+\alpha R^2$ as an exact dual formulation.  

Contrary to \eqref{risingStar}, the above \eqref{coefficients} does not allow for a simple analytic solution for a corresponding function $f(R)$ in the general case $\lambda/\mu\neq1/3$. However, just considering the leading term at large field values gives the differential equation
\begin{equation}
C_0 f'^2=\frac{f' R-f}{2f'}
\end{equation}
which is easily solved to yield
\begin{equation}
f(R)\sim R^{\thinspace 4/3}.
\end{equation}
This is an example for \eqref{genericn} and shows that even though the explicit function $f(R)$ is hard to find over the entire range of $R$, it must asymptote $R^{4/3}$ for large values.

Let us now investigate the limit of \eqref{wess} when the ratio $\lambda/\mu$ is perturbed around the value of $1/3$ such that 
\begin{equation}\label{deltaWess}
\lambda/\mu\rightarrow \frac{1}{3}-\delta,
\end{equation}
where $\delta$ is infinitesimally small. From \eqref{wess} we can infer the coefficients in \eqref{coefficients} to be
\begin{align}
C_0&=\frac{1}{4}\left[\frac{1}{4}-\frac{3\lambda}{2\mu}+\left(\frac{3\lambda}{2\mu}\right)^2 \right] \\
C_1&=\frac{1}{4}\left[3\frac{\lambda}{\mu}-\left(3\frac{\lambda}{\mu} \right)^2\right],
\end{align}
where the other $C_i$ are not of interest for reasons to become clear in a moment. Plugging \eqref{deltaWess} into the above, we find that\footnote{For $\lambda/\mu=1/3$, we have $C_0, C_1=0$. It is the fact that $C_0\sim\delta^2$ and $C_1\sim\delta$, in other words, $C_0$ approaches zero faster than $C_1$ which has \eqref{ratio} scale as $\delta$ even though $C_1(1/3)=0$.} 
\begin{equation}\label{ratio}
C_0/C_1\sim \delta.
\end{equation}
In other words, if $\lambda/\mu$ is perturbed with an infinitesimally small $\delta$, the squared rising exponential in \eqref{coefficients} is drastically suppressed with respect to the $C_1$-term. Thus one can well approximate \eqref{coefficients} by a scalar potential such as \eqref{risingStar} for which the corresponding $f(R)$-dual \eqref{risingF} is exactly known. 

This scenario \cite{Ellis2013} is not only a supergravity realisation of the vanilla $R^2$-inflation model, but also maintains the duality for an infinitesimal perturbation of the parameters of the model as in \eqref{deltaWess}. 

\subsection{Fibre Inflation}

The authors of \cite{0808.0691} develop a closed string inflation model where the inflaton field is a K\"ahler modulus with a string loop generated potential of the form
\begin{equation}\label{generalV}
V(\phi)=V'_0\left(\mathcal C_0 e^{\kappa'\phi}-\mathcal C_1e^{-\kappa'\phi/2}+\mathcal C_2e^{-2\kappa'\phi}+\mathcal C_{up} \right),
\end{equation}
with $\kappa'=2/\sqrt 3$. This potential maintains an approximately shift-symmetric plateau before a rising exponential starts to dominate. As the first negative exponential is the fourth root of the second and the coefficient in the exponent is larger than the Starobinsky $\kappa$, this model features an enhanced gravitational wave signal compared to the Starobinsky model, without reaching a tensor signal of comparable order of magnitude to that of chaotic inflation. In this section we are interested in investigating whether the fibre inflation potential has an approximate $f(R)$-description.

As before it is useful to recast \eqref{generalV} in terms of $\kappa=\sqrt{2/3}$.  In the low $\phi$ regime, where the rising exponential is negligible,  one obtains
\begin{equation}
V(\phi)=V_0 \left(1-\mathcal C_1e^{-{\frac{\kappa}{\sqrt{2}}}\phi}+\mathcal C_2e^{-2\sqrt{2}\kappa\phi}\right).
\end{equation}
We thus have
\begin{equation}
f(R)=\frac{1}{8V_0}R^2+\alpha\int f'^{1-1/\sqrt{2}}dR-\beta\int f'^{1-2\sqrt{2}}dR,
\end{equation}
which, upon enforcing the large $R$ regime\footnote{Here, we require large $R$, yet still small such the rising exponential has no effect.}, yields
\begin{equation}
f(R)=\frac{1}{8V_0}R^2+\alpha'R^{2-1/\sqrt{2}}+\beta'R^{2-2\sqrt{2}}+\ldots,
\end{equation}
where $\alpha', \beta'$ are rescaled coefficients due to the integration. We hence find an approximate $f(R)$-dual for the inflationary regime of the fibre inflation potential.

Considering now the regime in which the rising exponential dominates, we have the approximate potential
\begin{equation}
V(\phi)=V_0\thinspace\thinspace\mathcal C_{0}e^{\sqrt{2}\kappa\phi}.
\end{equation}
Using Eq. \eqref{potentialDef} we obtain
\begin{equation}
f'^{1+\sqrt{2}}\sim R.
\end{equation}
and therefore
\begin{equation}
f(R) \sim R^{\sqrt{2}}.
\end{equation}
This is in accordance with our previous findings and demonstrates that the leading order term in an $f(R)$-theory corresponding to a rising exponential has to be of type $R^{\thinspace n}$ with $1<n<2$. 

\subsection{Modified Fibre Inflation?}

One can ask whether it is possible to modify the above set-up such that the coefficients in the  exponents are the ones required to have integer powers of $f'$ in the differential equation one has to solve to find a corresponding $f(R)$ theory. We would then effectively have a situation such as \eqref{harddiff} which though not fully soluble, has separate  analytic solutions for both the regime in which the rising exponential dominates as well as the inflationary regime. These two solutions would be exact and the point of matching could in principle be determined. To answer that question, we have to take a closer look at the string construction of fibre inflation and see how it may be modified.

The reader looking for a review Fibre inflation is pointed towards the original work \cite{0808.0691} or the summary in \cite{1309.3413} as we will only recall here the bare minimum required for the present work. The crucial ingredient in the determination of the coefficient in the exponent is the normalisation of the kinetic term of the fibre modulus that drives inflation\footnote{We are excluding the possibility of manipulating the form of the string loop generated K\"ahler potential. Though this is possible in principle we consider it to be less well motivated than the modification of the volume form scrutinised here. }. The kinetic term itself derives from the volume of the compactification which, in the case of Fibre inflation, is given as
\begin{equation}\label{VolFibre}
\mathcal V=\alpha\left(\sqrt{\tau_1}\tau_2-\gamma\tau_3^{3/2} \right).
\end{equation}
This choice leads to $\kappa'=2/\sqrt{3}$, corresponding to the following relation between the fibre modulus $
\tau_1$ and the canonically normalised field $\phi$:  $\tau_1=e^{\kappa \phi}$. Equation \eqref{VolFibre} when combined with the conjectured form of the K\"ahler potential generated by string loops  on Calabi-Yau manifolds \cite{Cicoli:2007,Cicoli:2008} determines the potential \eqref{generalV} for the lightest K\"ahler modulus, $\tau_1$. As we have seen above this does not allow us to find an exact $f(R)$ formulation for the model.

One may however consider a slightly modified volume form, where the would be inflaton is still called $\tau_1$ but now corresponds to the volume of the base manifold rather that of the fibre as in the original setting. This amounts to considering 
\begin{equation}
\mathcal V=\alpha\left(\sqrt{\tau_2}\tau_1-\gamma\tau_3^{3/2} \right),
\end{equation}
which yields $\kappa'=1/\sqrt{3}$ and so once again one ends up with irrational powers of $f'$ and is therefore unable to solve the associated differential equation. 

Alternatively one may compactify on a torus, such that 
\begin{equation}
\mathcal V=\alpha\left(\sqrt{\tau_1\tau_2\tau_3}-\gamma\tau_3^{3/2} \right),
\end{equation}
yielding $\kappa'=1$. This exhausts the set of more obvious choices for $\mathcal V$ which has to be a polynomial of degree $3/2$ in the four-cycle volumes $\tau_i$. Whether or not Fibre inflation or a variation thereof can hence be modified in such a way as to maintain the Starobinsky $\kappa=\sqrt{2/3}$ is inconclusive as of now and thus more work seems to be required. 

\section{Power Suppression at low-$\ell$}\label{LowEll}

So far we have considered theoretical implications for a dual $f(R)$-description once a rising exponential has been added to the potential $V(\phi)$. We found that we either have to give up the dominating quadratic term, or that we have to limit the field range of $\phi$. We now turn our attention to a phenomenological fingerprint of a steepened potential $V(\phi)$. 

As discussed in various contexts~\cite{0303636,1211.1707,1309.3412,1309.3413,1309.4060,1404.2278,Kallosh:2014xwa}, a steepening of the inflaton potential in the vicinity of the point of 55 e-foldings can suppress power in the CMB temperature spectrum at large angular scales, i.e.\ low-$\ell$. 

At low multipoles, the power spectrum of the primordial curvature perturbation is given by \cite{09075424}
\begin{equation}\label{spectrum}
l(l+1)C^{TT}_l\propto\Delta^2_s(k) \propto \left(k/k_*\right)^{n_s-1},
\end{equation}
where $n_s=1+2\eta_V-6\epsilon_V$ is the spectral index, $\eta_V, \epsilon_V$ are the potential slow-roll parameters and $k_*$ is a pivot scale. Power loss is generated when the spectral index increases with decreasing wavenumber $k$. In order to obtain power suppression within the first few observable e-folds, we require $n_s$ to fall off sufficiently fast. Thus what remains is to investigate whether the corrected potentials mentioned above provide enough running of the spectral index. To do so, we evaluate $\Delta_s^2(k)$ at the onset of observable e-folds for the models \eqref{risingStar} and \eqref{rcubed} and compare the predictions with those of a spectrum with no running of $n_s$.

Considering potential \eqref{risingStar} or equivalently its $f(R)$ dual \eqref{risingF}, we find $n_s(55)=0.970$ and $n_s(62)=0.975$ and the power suppression is 1.7\% to 3.2\% where we have chosen $\delta\sim\mathcal O(10^{-4})$ to allow for a sufficient amount of e-folds and a red $n_s$. 

Considering scenario \eqref{rcubed}, we choose $\alpha\sim 10^{9}$, $|\beta|\sim2\cdot10^{14}$ to satisfy the boundary conditions for the number of e-folds and a red $n_s$. The spectral index takes similar values as for \eqref{risingStar} and the power suppression is 1.8\% to 3.3\%. Lowering the amount of observable e-folds to $N_{obs}\sim 50$ by assuming intermediate reheating temperatures and adjusting $|\beta|\rightarrow 10^{15}$ yields $n_s(45)= 0.978$ and $n_s(50)= 0.985$ and thus gives a suppression of about 5\%.

\section{Discussion}\label{concl}

In this letter, we investigated the duality between scalar field theories being coupled minimally and non-minimally to gravity and $f(R)$-Lagrangians.

We showed that any potential with an infinitely long plateau may be recast as an $f(R)$-theory which is $\sim R^2$ to leading order. We further demonstrated how to obtain an expression for an $f(R)$-Lagrangian driving chaotic inflation within the accessible range of inflationary e-folds. Weyl-rescaling from the $f(R)$-frame to another Jordan frame, we found a general expression relating an arbitrary $f(R)$-theory to a scalar field theory non-minimally coupled to gravity and consequently established that a series of logarithmic corrections in any Jordan frame leads to chaotic inflationary dynamics as becomes apparent when analysing the resulting Einstein frame inflaton potential. Having noted the different dynamics of the inflationary spacetime regarding the cut-off of the series of logarithmic corrections, we learn that higher order terms of the Ricci scalar are of crucial importance and may not be neglected if they exist.

Turning to modifications of plateau-like potentials at higher field values, we gave the example of potential \eqref{fibre} which looses its {closed form} dual $f(R)$ description once a rising correction is added. We showed that any scalar potential which is dominated by rising exponentials at higher field values at least allows for an $f(R)$-dual that is $\sim R^n$ with $1<n<2$ to leading order. The Starobinsky potential \eqref{staro} maintains a {closed form} dual formulation \eqref{risingF} when a rising exponential is considered \eqref{risingStar}. The important implication however is that the leading $R^2$-term is removed from the $f(R)$-dual, as expected. The resulting $f(R)$ theory is an infinite series which sums to an expression which is to leading order $R^{3/2}$. Furthermore, the correction may not come with arbitrary powers of exponential functions as \eqref{differentialequation} has to remain soluble to find an analytic expression for the function $f(R)$. Thus the form of correction to the Starobinsky potential is strongly limited if the duality shall be given by explicit expressions obtainable in closed form by known methods over the entire range. Only considering the limit of large field values, approximate $f(R)$ duals may easily be found explicitly. {It is important to note that the inability to find closed form $f(R)$ duals does not exclude the existence of a dual description in terms of a theory of modified gravity. It merely demonstrates that the dual description is out of our reach with current methods.}

Concluding, the inflationary behaviour and predictions can be substantially changed by considering slight corrections to the first two powers in a polynomial $f(R)$-theory. Investigating only finite order corrections, we pay the price of either having to place the inflaton on the right side of a hilltop or having a finite field range for the inflaton $\phi$, depending on the sign of the higher power's coefficient. 

Therefore, corrections to the Einstein frame potential with rising exponentials - provided an $f(R)$ dual may be found - remove the leading $R^2$-term whereas higher powers in $R$ with negative coefficients limit the field range. Both are interesting theoretical consequences. 

Considering concrete UV examples, we find that scenario \eqref{wess} can maintain an $f(R)$-dual \eqref{risingF} given a certain choice of model parameters. The string inflation scenario \eqref{generalV} does not allow for an exact $f(R)$-dual. Modifications of the string theory set-up, i.e.\ considering different Calabi-Yau compactifications, did not prove to be successful at a first attempt. 

Phenomenologically, rising exponentials coming to dominate the potential at higher field values induce some running of the spectral index $n_s$. However, considering \eqref{risingF} the running is small and the consequent effect of power suppression at low-$\ell$ is up to $\sim 3\%$. Hence if the observational significance of power suppression increases and the effect is found to be $\gtrsim 3\%$, one requires corrections to an Einstein frame potential of higher order than those that can be provided by an exact dual $f(R)$ description. Thus in case of no significant detection of inflationary tensor modes, a significant power-loss at low-$\ell$ would present another route to the disfavouring of $R^2$-type inflation models. \\

\textbf{Acknowledgements.} \\
BB would like to thank Diederik Roest for insightful comments while this work was in preparation.
BB and AW are supported by the Impuls und Vernetzungsfond of the Helmholtz Association of German Research Centres under grant HZ-NG-603. This work has been supported by the ERC Advanced Grant SPLE under contract ERC-2012-ADG-20120216-320421 and by  the grants FPA2012-32828, and FPA 2010-20807-C02.
FGP would also like to thank the support of the spanish MINECO {\it Centro de excelencia Severo Ochoa Program} under grant SEV-2012-0249.


\appendix
\section{The $f(R)$ dual for $V\sim V_0$}\label{a1}

Recall the potentials $V=V_0(1-2e^{-\frac{\kappa}{n}\phi}+e^{-2\kappa\phi})$ with $n>1$ and $V=V_0(1-2e^{-\kappa\phi}+e^{-n\kappa\phi})$ with $n>2$. Both potentials have the exponentials departing from their square relation characteristic of the $R^2$ dual Starobinsky potential. The aim of this appendix is to prove the claim made in section 2, namely that regardless of the specific values chosen for $n$, both potentials will always admit at least an approximate $f(R)$ dual which is to leading order $R^2$. Essentially, one may argue that the potential mimics a cosmological constant for large field values and hence all one has to do is finding the $f(R)$ dual to general relativity with a free scalar and a cosmological constant. To that extent, consider that both potentials display a shift symmetry in the inflationary region, i.e.\ one may well approximate both of the above as
\begin{equation}\label{a1}
V(\phi)\sim V_0
\end{equation}
during inflation. We recall equation \eqref{potentialDef} and hence write
\begin{equation}\label{b2}
V_0=\frac{f'\thinspace R-f}{2f'^2},
\end{equation}
which, upon rearranging, may be recast as
\begin{equation}
2V_0f'^2-f'\thinspace R+f=0.
\end{equation}
Differentiating the above with respect to $R$ gives, for $f''\neq 0$,
\begin{equation}
4V_0f'-R=0,
\end{equation}
which may then simply be integrated to yield
\begin{equation}\label{a55}
f(R)=\frac{1}{8V_0}R^2.
\end{equation}
The integration constant has been set to zero by considering the boundary condition \eqref{b2}. We thus see that any potential which approximates a cosmological constant, i.e.\ $V\sim V_0$ , may be recast in terms of a leading order $R^2$ $f(R)$ formulation. The same argument also applies vice-versa, i.e.\ the scale invariant theory $f(R)=\alpha R^2$ may readily be recast in terms of an Einstein-Hilbert Lagrangian with a cosmological constant $\Lambda=(8\alpha)^{-1}$.

When considering e.g.\ the full potential $V(\phi)=V_0(1-2e^{-\frac{\kappa}{n}\phi}+e^{-2\kappa\phi})$ with $n>1$, one has, according to expression \eqref{potentialDef},
\begin{equation}
\frac{f' R-f}{2f'^2}=V_0\left(1-2f'^{-\frac{1}{n}}+f'^{-2}\right),
\end{equation}
where we have identified $f'=e^{\kappa\phi}$. Rearranging and differentiating with respect to $R$ gives
\begin{equation}\label{a7}
f'=\frac{R}{4V_0}+\left(2-\frac{1}{n}\right)f'^{1-\frac{1}{n}} .
\end{equation}
For large values of $R$, we hence approximate $f'\sim \mathcal O(R)$. Therefore, we plug $f'\sim R$ back into \eqref{a7} to obtain
\begin{equation}
f(R)=\frac{R^2}{8V_0}+R^{2-\frac{1}{n}}+\ldots
\end{equation}
where the dots denote sub-leading terms during inflation and indicate that the above was obtained iteratively. We thus find that the leading order behaviour is indeed $R^2$. Applying the same procedure to the potential $V=V_0(1-2e^{-\kappa\phi}+e^{-n\kappa\phi})$ with $n>2$ yields as a first iterative step
\begin{equation}
f(R)=\frac{R^2}{8V_0}+\frac{1}{2}R^{2-n}+\ldots
\end{equation}
where again the dots denote sub-leading terms. For the above, one might wonder whether or not the function $f(R)$ becomes singular for small $R$. However, the above was explicitly obtained with a method relying on limiting the range of validity of the solution to the large R regime. Thus it is found that the enhanced $R^2$ term dominates, as was foreseen from expressions \eqref{a1} to \eqref{a55}.

\section{An Explicit Derivation}\label{a2}

Recall the potential
\begin{equation}\label{p1}
V(\phi)=V_0\left(1-e^{-\gamma\kappa\phi} \right)^2,
\end{equation}
with $\kappa=\sqrt{2/3}$ and $0<\gamma<2$. By considering \eqref{potentialDef}, we write
\begin{equation}
V_0\left(1-2e^{-\gamma\kappa\phi}+e^{-2\gamma\kappa\phi} \right)=\frac{f'\thinspace R-f}{2f'^2},
\end{equation}
which, upon identifying $f'=e^{\kappa\phi}$ and multiplying by $2f'^2$, can be rewritten as
\begin{equation}
2V_0\left(f'^2-2f'^{2-\gamma}+f'^{2-2\gamma}\right)+f-f'\thinspace R=0.
\end{equation}
Differentiating with respect to $R$ yields
\begin{equation}
	2V_0\left[2f'-2(2-\gamma)f'^{1-\gamma}+(2-2\gamma)f'^{1-2\gamma}\right]- R=0,
\end{equation}
where, as always, we are taking $f''\neq 0$. Having reduced the rank of the differential equation, we may now rearrange terms and integrate to write
\begin{equation}
f(R)=\frac{1}{8V_0}R^2+(2-\gamma)\int f'^{1-\gamma}dR \thinspace-(1-\gamma)\int f'^{1-2\gamma}\thinspace dR.
\end{equation}
This establishes that for large $R$, $f'(R)$ may be approximated as being of order $\sim R$. We hence state the approximate $f(R)$ dual to potential \eqref{p1} as
\begin{equation}
f(R)=\frac{1}{8V_0}R^2 + R^{2-\gamma} - \frac{1}{2}R^{2-2\gamma} +\ldots
\end{equation}
up to sub-leading terms where the above solution may be understood as the first result of an iterative approach. Connecting to the expressions in the text, we thus identify
\begin{equation}
\alpha=2-\gamma,\quad \beta=\gamma-1,\quad c=1.
\end{equation}

\bibliography{r3_13}

\end{document}